\documentstyle[aps, preprint]{revtex}
\begin{document}
\draft
\title{The general double-dust solution}
\author{B.V.Ivanov\thanks{%
E-mail: boyko@inrne.bas.bg}}
\address{Institute for Nuclear Research and Nuclear Energy,\\
Tzarigradsko Shausse 72, Sofia 1784, Bulgaria}
\maketitle

\begin{abstract}
The gravitational field of two identical rotating and counter-moving dust
beams is found in full generality. The solution depends on an almost
arbitrary function and a free parameter. Some of its properties are studied.
Previous particular solutions are derived as subcases.
\end{abstract}


\section{Introduction}

The study of the gravitational field of light beams (null dust) has a long
history. In the linear approximation to general relativity this was done 71
years ago \cite{one,two}. Later, Bonnor found exact solutions which belong
to the class of pp-waves (algebraically special solutions of Petrov type N
and no expansion) \cite{three,four}. He showed that beams shining in the
same direction do not interact. The gravitational field of two
counter-moving light beams is very complicated and particular solutions were
found only recently \cite{five,six}. One of them was generalized to the case
of two colliding non-null dusts and named the double-dust solution \cite
{seven}. It is based on a one-parameter relation between two metric
components.

In this paper we find the general double-dust solution and study its
properties. It depends on one almost arbitrary function and one free
parameter. All previous solutions are derived as special cases.

In section 2 the field equations are written and their general solution is
given in three different ways. The condition of elementary flatness is
imposed, which reduces the number of free parameters to one. The
restrictions on the arbitrary function, the positivity of the energy density
and the regular character of the metric are studied. In section 3 some
particular examples are given such as the solutions of Lanczos \cite{eight},
Lewis \cite{nine} and the two Kramer solutions (for null and non-null dust).
In section 4 the properties of the 4-velocities of the two dust beams are
studied. It is proved that the general solution satisfies the dominant
energy condition. The appearance of closed timelike curves is discussed in
the stationary version of the Kramer double-dust solution. In section 5 the
general interior solution is matched smoothly to the vacuum Lewis solution
at any distance from the axis. Section 6 contains a short discussion.

\section{Field equations and the general solution}

We shall work in the stationary formalism when the cylindrically symmetric
metric is given by \cite{four} 
\begin{equation}
ds^2=-e^{2u}\left( dt+Ad\varphi \right) ^2+e^{-2u}\left[ e^{2k}\left(
dx^2+dz^2\right) +W^2d\varphi ^2\right]  \label{one}
\end{equation}
and there is only radial $x$-dependence. The static metric may be obtained
by complex substitution: $t\rightarrow iz$, $z\rightarrow it$, $A\rightarrow
iA$. The energy-momentum tensor of the two dust beams reads 
\begin{equation}
T_{\mu \nu }=\mu _1k_\mu k_\nu +\mu _2l_\mu l_\nu ,  \label{two}
\end{equation}
\begin{equation}
k_\mu k^\mu =-1,\qquad l_\mu l^\mu =-1  \label{three}
\end{equation}
where $\mu _i$ are the density profiles of the rotating and flowing beams.
The only components of their 4-velocities are $k^t=l^t=V$, $k^z=-l^z=Z$. Eq.
(3) becomes 
\begin{equation}
e^{2u}V^2=1+e^{2\left( k-u\right) }Z^2.  \label{four}
\end{equation}

To write the Einstein equations we use the combinations of Ricci tensor
components utilized in Refs. \cite{ten,eleven,twelve,thirteen}. We shall
work, however, with curved mixed components $R_\nu ^\mu $. The only
non-trivial ones are the diagonal components and $R_\varphi ^t$, $%
R_t^\varphi $. The Einstein equations for the corresponding $T_\nu ^\mu $
components give after some rearrangement Eq. (4) and 
\begin{equation}
k^{\prime \prime }=-u^{\prime 2}-B,\qquad B=\frac{a_0^2}4e^{-4u},
\label{five}
\end{equation}
\begin{equation}
k^{\prime }\frac{W^{\prime }}W=u^{\prime 2}-B,  \label{six}
\end{equation}
\begin{equation}
A^{\prime }=a_0We^{-4u},  \label{seven}
\end{equation}
\begin{equation}
8\pi \mu e^{2\left( k-u\right) }=u^{\prime \prime }+\frac{u^{\prime
}W^{\prime }}W-\frac{W^{\prime \prime }}W+2B,  \label{eight}
\end{equation}
\begin{equation}
16\pi \mu Z^2e^{4\left( k-u\right) }=\frac{W^{\prime \prime }}W.
\label{nine}
\end{equation}
Here $^{\prime }$ is a $x$-derivative, $2\mu =\mu _1+\mu _2$ and $a_0$ is an
integration constant resulting from Eq. (7). Units are used with $G=c=1$.
The equations for the rest $T_\nu ^\mu $ components are satisfied either
identically or when $\left( \mu _1-\mu _2\right) Z=0$. Therefore, we accept
in the following the relation $\mu _1=\mu _2=\mu $. In the case of a single
beam, we must put instead $Z=0$, which halts its motion along the axis.

There are 6 equations (4-9) for 7 quantities; $u,k,W,A,\mu ,Z,V$. Let us
take $u\left( x\right) $ to be an arbitrary function. The equations decouple
and starting with Eq. (5) and finishing with Eq. (4) the unknowns are found
in the above-mentioned order. The function $k$ is obtained by integrating
Eq. (5). Thus we have 
\begin{equation}
k^{\prime }\left( x\right) =k_uu^{\prime }\left( x\right) =-\int_0^x\left(
u^{\prime 2}+B\right) dx  \label{ten}
\end{equation}
and $k_u$ may be expressed either as a function of $x$ or $u$ if $u^{\prime
}\left( x\right) $ or $u^{\prime }\left( u\right) $ is used. The other
quantities also have double representations 
\begin{equation}
W=W_0k^{\prime }e^{2\int_0^x\frac{u^{\prime 2}}{k^{\prime }}%
dx}=W_0k_uu^{\prime }e^{2\int_0^u\frac{du}{k_u}},  \label{eleven}
\end{equation}

\begin{equation}
A=a_0W_0\int_0^xk^{\prime }e^{2\int_0^x\frac{u^{\prime 2}}{k^{\prime }}d\bar 
x-4u}dx=a_0W_0\int_0^uk_ue^{2\int_0^u\frac{d\tilde u}{k_{\tilde u}}-4u}du,
\label{twelve}
\end{equation}
\begin{eqnarray}
8\pi \mu  &=&\left( 1-\frac{2u^{\prime }}{k^{\prime }}\right) \left(
u^{\prime \prime }+\frac{u^{\prime 3}}{k^{\prime }}+\frac{2k^{\prime
}-u^{\prime }}{4k^{\prime }}a_0^2e^{-4u}\right) e^{-2\left( k-u\right) }= 
\nonumber  \label{thirteen} \\
&&\frac{k_u-2}{k_u^2}\left[ \frac{a_0^2}2\left( k_u-1\right)
e^{-4u}-k_{uu}u^{\prime 2}\right] e^{-2\left( k-u\right) },  \label{thirteen}
\end{eqnarray}
\begin{equation}
Z^2=\frac{u^{\prime }e^{-2\left( k-u\right) }}{k^{\prime }-2u^{\prime }}=%
\frac 1{k_u-2}e^{-2\left( k-u\right) },  \label{fourteen}
\end{equation}
\begin{equation}
V^2=\frac{k^{\prime }-u^{\prime }}{k^{\prime }-2u^{\prime }}e^{-2u}=\frac{%
k_u-1}{k_u-2}e^{-2u}.  \label{fifteen}
\end{equation}
Eq. (14) shows that $k_u>2$. The second representation requires the passage
to a new radial variable $u$ or $f=e^u$ and correspondingly $g_{xx}$ becomes 
\begin{equation}
g_{uu}=\frac{e^{2\left( k-u\right) }}{u^{\prime 2}},  \label{sixteen}
\end{equation}
where $u^{\prime }=u^{\prime }\left( u\right) $.

One can pass to $u$ in another way, by defining the arbitrary function $%
g\left( u\right) =k^{\prime }$. Then Eq. (5) becomes a quadratic equation
for $u^{\prime }$ with solutions 
\begin{equation}
2h\left( u\right) =2u^{\prime }=-g_u\pm \sqrt{g_u^2-4B}  \label{seventeen}
\end{equation}
and $k_u=g/h$. Eqs. (11)-(15) are easily rewritten in terms of $g\left(
u\right) $. The restriction $k_u>2$ holds iff 
\begin{equation}
g=\sqrt{e^{-u}F\left( \lambda \right) },\qquad \lambda =e^{-3u},
\label{eighteen}
\end{equation}
where $F$ is an arbitrary positive function, satisfying $F_\lambda >a_0^2/3$%
. A sufficient condition for the radical in Eq. (17) to be real is

\begin{equation}
F>2a_0^2\lambda ,  \label{nineteen}
\end{equation}
which encompasses the previous inequality.

In the above equations we have set $u\left( 0\right) =0$, so the axis of
rotation still coincides with the $z$-axis. The indefinite integral in Eq.
(11) was replaced by a definite one and the integration constant $W_0$.
Another constant $k\left( 0\right) =k_0$ appears when Eq. (10) is further
integrated. The condition for elementary flatness 
\begin{equation}
\lim\limits_{x\rightarrow 0}\frac{e^{u-k}}x\left(
e^{-2u}W^2-e^{2u}A^2\right) ^{1/2}=1  \label{twenty}
\end{equation}
must be satisfied at the axis. $A$ should vanish there which is accounted
for by the lower limit of the integral in Eq. (12). $W$ should vanish too.
One can see from Eq. (7) that $A=o\left( W\right) $ when $x\rightarrow 0$
and can be neglected in Eq. (20) like in many other cases \cite
{eleven,twelve,thirteen}. Therefore we obtain the static case condition 
\begin{equation}
\lim\limits_{x\rightarrow 0}\frac{e^{-k}\left| W\right| }x=1.
\label{twentyone}
\end{equation}
We inset the expression for $W$ from Eq. (11) and take the limit. Eq. (10)
gives 
\begin{equation}
k_u\left( 0\right) =-\frac{a_0^2}{4u^{\prime \prime }\left( 0\right) },
\label{twentytwo}
\end{equation}
so that $u^{\prime \prime }\left( 0\right) <0$ and is finite. Then Eq. (11)
shows that $W\left( 0\right) =0$ due to $u^{\prime }\left( 0\right) =0$. Eq.
(21) yields 
\begin{equation}
a_0^2\left| W_0\right| =4e^{k_0}.  \label{twentythree}
\end{equation}
Hence, elementary flatness gives a relation between the 3 integration
constants without affecting the arbitrary functions. The constant $k_0$ may
be put zero by a coordinate change. We won't do this for matching purposes.
Thus the general solution depends on one free parameter and the function $%
u\left( x\right) $, $u^{\prime }\left( u\right) $ or $g\left( u\right) $.
The only restriction for $u^{\prime }\left( u\right) $ comes from the
condition $k_u>2$. For $g\left( u\right) $ this leads to Eq. (19), while $%
u\left( x\right) $ should satisfy 
\begin{equation}
-\int_0^x\left( u^{\prime 2}+B\right) dx>2u^{\prime }.  \label{twentyfour}
\end{equation}
Both sides vanish on the axis, hence, a sufficient condition is 
\begin{equation}
2u^{\prime \prime }+u^{\prime 2}+B<0\qquad \text{or}\qquad -f^{7/2}\left(
f^{1/2}\right) ^{\prime \prime }>\frac{a_0^2}{16}  \label{twentyfive}
\end{equation}

An important physical requirement is $\mu >0$ in some region around the
axis, which is enough for a interior solution. According to Eq. (13) $\mu
\left( 0\right) \geq 0$ always. If $k_{uu}<0$, then $\mu $ is positive
everywhere. Otherwise the energy density may become negative at some
distance from the axis.

Eq. (16) indicates that the change of coordinates $x\rightarrow u$ is always
singular at the origin. This may be avoided by introducing a third radial
coordinate $r$, such that $u=u\left( r^2\right) $ \cite{seven}. This is
equivalent to going back to the first version of the solution by choosing $%
u\left( x\right) $. However, some solutions look simpler when $u$ is the
radial coordinate.

\section{Some particular solutions}

Let us derive several concrete solutions. For a single beam we must set $Z=0$
in order to use metric (1) and the field equations above. Then Eq. (14)
gives constant $u$, which is set to zero. The solution should be found from
Eqs. (4-9). We obtain 
\begin{equation}
V=1,\qquad W=x,\qquad A=\frac 12a_0x^2,  \label{twentysix}
\end{equation}
\begin{equation}
k=-\frac 18a_0^2x^2,\qquad 8\pi \mu =\frac 12a_0^2e^{-2k}.
\label{twentyseven}
\end{equation}
This is the cherished Lanczos solution \cite{eight} in comoving coordinates.
It represents a cylinder of rigidly rotating dust. In the absence of $%
u\left( x\right) $ it depends just on the parameter $a_0$.

Another important case is $\mu =0$, which, at first sight, should lead to
vacuum solutions. Two obvious candidates are $k=2u$ and $k=u$. Both of them
belong to the one-parameter series $k=\left( a+1\right) u$. This case was
solved \cite{seven} in the static formulation of the problem. It is worth to
do this in the stationary frame and see the differences. We must have $a>1$
and $k_0=0$. Eqs. (14-15) give directly 
\begin{equation}
Z^2=\frac{f^{-2a}}{a-1},\qquad V^2=\frac{af^{-2}}{a-1}.  \label{twentyeight}
\end{equation}
Eqs. (11,12) yield the expressions 
\begin{equation}
W=\left( a+1\right) W_0f^{\prime }f^{\frac{1-a}{1+a}},  \label{twentynine}
\end{equation}
\begin{equation}
A=\frac{a_0W_0\left( a+1\right) ^2}{2\left( 2a+1\right) }\left( 1-f^{-2\frac{%
2a+1}{a+1}}\right) ,  \label{thirty}
\end{equation}
Eq. (23) gives $W_0=4/a_0^2$. It is necessary to express $f^{\prime }$
through $f$. Eqs. (5-6) yield 
\begin{equation}
k^{\prime }=-\frac{a_0A}{2W}.  \label{thirtyone}
\end{equation}
Combining the last three equations gives the result 
\begin{equation}
f^{\prime 2}=\frac{a_0^2}{4\left( 2a+1\right) }\left( f^{-2}-f^{\frac{2a}{a+1%
}}\right) .  \label{thirtytwo}
\end{equation}
Positivity requires $u<0,f<1,k<0$. This equation may be integrated 
\begin{equation}
\frac{a_0\sqrt{2a+1}}{a+1}x=\frac{\Gamma \left( -\frac 1{2\left( 2a+1\right) 
}\right) \Gamma \left( \frac 12\right) }{\Gamma \left( \frac a{2a+1}\right) }%
-B_\xi \left( -\frac 1{2\left( 2a+1\right) },\frac 12\right) ,
\label{thirtythree}
\end{equation}
where $\xi =f^{-2\frac{2a+1}{a+1}}$ and $B_\xi \left( p,q\right) $ is the
incomplete beta function. The formula above can not be inverted, but this is
not necessary, since we have chosen a new radial variable. Inserting Eq.
(32) into Eqs. (29), (16) and (13) we find explicit expressions for $%
W,g_{uu} $ and $\mu $, which simplifies to 
\begin{equation}
8\pi \mu =\frac{a_0^2a\left( a-1\right) }{2\left( a+1\right) ^2}f^{-2a-4}.
\label{thirtyfour}
\end{equation}
The gravitational field is determined completely.

In the static formulation $0\leq a<1,f>1,k>0$. Here we have the opposite.
The sign of $a-1$ is changed correspondingly. The model has two parameters, $%
a$ (remnant of the arbitrary function) and $a_0$. Kramer puts $a_0=1$. Then
the central density is bounded, $8\pi \mu \left( 0\right) \leq 1/2$. In
fact, it can be as big as we want it, as seen from Eq. (34). In principle,
one can choose an arbitrary density profile, but the extraction of $g\left(
u\right) $ out of it is not possible explicitly. In Ref. \cite{seven} it is
asserted that $f\left( x\right) $ has no analytical expression. This is
true, however, Eq. (33) shows that $x\left( f\right) $ does have one in
special functions.

Let us look at the case $a=1$. The factors $a-1$ in $\mu ,Z^2,V^2$ cancel
each other, while Eq. (4) becomes $V^2=Z^2$. The two beams turn into
null-dust and the metric coincides with Eq. (21) (with $\lambda =\sqrt{3}/4$%
) from Ref. \cite{five} when $a-1$ is set effectively to $2$, $a_0=1$ and $%
\xi =1+3r^2/16$. This is the proper choice of the final radial coordinate $r$%
, instead of $\xi =\left( 1+r^2\right) ^{-1}$ in the static case.

The case $a=0$ is a true vacuum case. The static metric becomes simply flat
spacetime. Here we have 
\begin{equation}
W=x,\qquad f=x^2-c_0^2,\qquad A=-\frac{a_0x^2}{2c_0^2f},\qquad e^{k-u}=1,
\label{thirtyfive}
\end{equation}
with $c_0$ being an integration constant. We should compare this metric to
the general Lewis solution \cite{nine,fourteen,fifteen} 
\begin{equation}
W_L=x,\qquad f_L=lx^{1-n}-\frac{c^2}{n^2l}x^{n+1},  \label{thirtysix}
\end{equation}
\begin{equation}
A_L=\frac{cx^{n+1}}{nlf_L}+b,\qquad e^{2\left( k_L-u_L\right) }=x^{\frac 12%
\left( n^2-1\right) },  \label{thirtyseven}
\end{equation}
which includes 4 constants $b,c,l,n$. Obviously, Eq. (35) is a particular
case of Eqs. (36,37) with $n=-1$, $l=1$, $c=c_0$, $b=-c_0^{-1}$ and $%
2c_0=a_0 $. The solution belongs to the Weyl class and has a regular
one-parameter metric when $x^2>c_0^2$.

\section{Properties of the general solution}

Let us study first the properties of $k^\mu $ and $l^\mu $. The 4-velocities
are geodesic (no acceleration) and have zero expansion and shear. The
vorticity (twist) vector $w^\mu $ has been calculated with the help of
GRTensor for the general cylindrical metric (1) and for the general
solution. We have $w^x=0$ and for $k^\mu $%
\begin{equation}
w^z=\frac{A^{\prime }V^2}{2W}e^{6u-2k}=\frac{a_0\left( k_u-1\right) }{%
2\left( k_u-2\right) }e^{-2k},  \label{thirtyeight}
\end{equation}
\begin{equation}
w^\varphi =-\frac{e^{2u}}{2W}\left[ 2VZ\left( 2u^{\prime }-k^{\prime
}\right) +ZV^{\prime }-Z^{\prime }V\right] =\frac{2\left( k_u-2\right)
\left( k_u-1\right) -k_{uu}}{4W_0k_u\sqrt{k_u-1}\left( k_u-2\right) }%
e^{2u-k-2\int_0^u\frac{du}{k_u}},  \label{thirtynine}
\end{equation}
\begin{equation}
w^t=-Aw^\varphi +\frac{A^{\prime }ZV}{2W}e^{2u}=-Aw^\varphi +\frac{a_0\sqrt{%
k_u-1}}{2\left( k_u-2\right) }e^{-2u-k}.  \label{forty}
\end{equation}
For $l^\mu $, $w^\varphi $ and $w^t$ change sign. These expressions coincide
in the Kramer case $k=\left( a+1\right) u$ with Eq. (39) from Ref. \cite
{seven} after the passage to the static metric is done and one takes into
account the difference between $x$ and $\zeta =f^2$.

Another useful quantity is $\Gamma =k^\mu l_\mu $. It reads for the
different metrics 
\begin{equation}
\Gamma =-\left[ 1+2e^{2\left( k-u\right) }Z^2\right] =-\frac{k_u}{k_u-2}=-%
\frac{a+1}{a-1}.  \label{fortyone}
\end{equation}
The proof of the dominant energy condition in Ref. \cite{seven} may be
lifted to the general solution. It depends crucially on the fact that $%
\Gamma ^2>1$, which is obvious from the above formula.

It is well known that closed timelike curves (CTC) exist when $g_{\varphi
\varphi }<0$. This condition is rather intractable further for the general
solution. In the Kramer case we have 
\begin{equation}
g_{\varphi \varphi }=\frac{4\left( a+1\right) ^2}{\left( 2a+1\right) ^2}%
f^{-2}\left[ \left( 2a^2+2a+1\right) f^{\frac 2{a+1}}-a^2f^{-\frac{4a}{a+1}%
}-\left( a+1\right) ^2f^4\right] .  \label{fortytwo}
\end{equation}
The sign is determined basically by the competition of the first and the
second terms in the square brackets. It becomes negative when 
\begin{equation}
\xi >1+\frac{a^2}{\left( a+1\right) ^2}.  \label{fortythree}
\end{equation}
This happens always for big enough $r$. In the static metric formulation $%
g_{\varphi \varphi }$ is strictly positive and there are no CTC as seen from
Ref. \cite{seven}, Eq. (36). In this case a matching can be done at some $%
r_0 $ to the Levi-Civita static metric and a realistic global solution
constructed. In the stationary case the matching should be done to the Lewis
solution (36,37) whose Weyl class is essentially static, but the Lewis class
contains CTC.

\section{The global solution}

We shall show that the general interior solution can be matched smoothly to
the Lewis solution (36,37) for any value $x_0$ of the radial coordinate. For
this purpose we perform scaling on two of the coordinates in the exterior, $%
\bar t=\tau t$ and $\bar \varphi =\Omega \varphi $. Instead of scaling $z$
we introduce again $k_0$ in the interior. In Ref. \cite{seven} the
transformation $f^2=h\left( \rho \right) $ was also applied, where $h$ is an
arbitrary function of the exterior radial coordinate $\rho $. The general
solution, however, already includes an arbitrary (modulo some positivity
requirements) function $u\left( x\right) $ and it is redundant to introduce
a second one. The effect of such transformation then is that $u\left(
x_0\right) $ and $u^{\prime }\left( x_0\right) $ become free parameters,
while quantities given by integrals, like $k\left( x_0\right) $, are fixed
by the interior solution. The matching procedure is the following: we choose
an interior solution which fixes $a_0$, $k\left( x_0\right) $, $k^{\prime
}\left( x_0\right) $, $W\left( x_0\right) ,$ $A\left( x_0\right) $ and $%
A^{\prime }\left( x_0\right) $ up to $W_0$. Next we demand that the metric
should be continuous together with its first derivative at the junction.
This gives several algebraic equations for the free constants $u\left(
x_0\right) $, $u^{\prime }\left( x_0\right) $, $k_0$, $\tau $, $\Omega $, $n$%
, $l$, $c$ and $b$. Finally, we fix them by solving the equations, the only
free parameter remaining being $x_0$. We consider the first representation
of the general solution and all functions in the following are taken at the
point $x_0$.

The continuity of the metric yields four conditions 
\begin{equation}
f_L^2\tau ^2=f^2,\qquad f_L^2A_L\tau \Omega =f^2A,  \label{fortyfour}
\end{equation}
\begin{equation}
\left( f_L^2A_L^2+f_L^{-2}W_L^2\right) \Omega ^2=f^2A^2+f^{-2}W^2,
\label{fortyfive}
\end{equation}
\begin{equation}
f_L^{-2}e^{2k_L}=f^{-2}e^{2k+2k_0}.  \label{fortysix}
\end{equation}
The continuity of the metric derivatives supplies another four. After some
rearrangement, the total system of 8 equations becomes 
\begin{equation}
e^{2k_0}=x_0^{\frac 12\left( n^2-1\right) }e^{2u-2k},  \label{fortyseven}
\end{equation}
\begin{equation}
n^2-1=4x_0\left( k^{\prime }-u^{\prime }\right) ,  \label{fortyeight}
\end{equation}
\begin{equation}
A_L=\frac{A^{\prime }}AA_L^{\prime },  \label{fortynine}
\end{equation}
\begin{equation}
\frac \Omega \tau =\frac A{A_L},\qquad \tau \Omega =\frac W{x_0},
\label{fifty}
\end{equation}
\begin{equation}
\frac{W^{\prime }}Wx_0=1,  \label{fiftyone}
\end{equation}
\begin{equation}
f_L^2A_L^{\prime }=a_0x_0f^{-2},  \label{fiftytwo}
\end{equation}
\begin{equation}
\frac{f_L^{\prime }}{f_L}=u^{\prime }.  \label{fiftythree}
\end{equation}
Eq. (47) determines $k_0$. Then Eq. (23) determines $W_0$ and, hence, $W$.
Eq. (48) determines $n$, while Eqs. (37,49) fix $b$. Eq. (50) yields
expressions for $\Omega $ and $\tau $%
\begin{equation}
\Omega ^2=\frac{a_0W^2}{x_0A_L^{\prime }f^4},\qquad \tau ^2=\frac{%
A_L^{\prime }f^4}{x_0a_0}.  \label{fiftyfour}
\end{equation}
Eqs. (51-53) form a system for $f$, $u^{\prime }$ and $c/l$%
\begin{equation}
u^{\prime 2}=\frac{4k^{\prime }f^8+x_0a_0^2}{4x_0f^8},  \label{fiftyfive}
\end{equation}
\begin{equation}
\frac cl=\frac{na_0x_0^{1-n}}{f^2\left( 1+n-2u^{\prime }x_0\right) },
\label{fiftysix}
\end{equation}
\begin{equation}
1-n-2x_0u^{\prime }=\frac{ca_0x_0^{n+1}}{nlf^2}.  \label{fiftyseven}
\end{equation}
Replacing Eq. (56) into Eq. (57) and using Eq. (48) we obtain 
\begin{equation}
4x_0^2u^{\prime 2}=x_0^2a_0^2f^{-4}+4x_0k^{\prime }.  \label{fiftyeight}
\end{equation}
The combination of Eqs. (55,58) gives 
\begin{equation}
f=1,\qquad u^{\prime 2}=\frac{4k^{\prime }+x_0a_0^2}{4x_0}.
\label{fiftynine}
\end{equation}
Inserting these formulas in the previous equations, everything is determined
in terms of the interior solution and $x_0$. For example, we have for the
parameter $n$, which measures the line mass density 
\begin{equation}
n^2=1+4x_0k^{\prime }-\left[ 4x_0\left( 4k^{\prime }+x_0a_0^2\right) \right]
^{1/2}.  \label{sixty}
\end{equation}
There are several interesting features of the matching. The gravitational
potential $f$ does not depend on any parameters at the junction and has the
same value as at the axis. Only the ratio of $l$ and $c$ is determined. The
quantity $n^2$ is not necessarily positive. When $n^2<0$ we enter the Lewis
class of the Lewis solution, which possesses CTC \cite{fourteen,fifteen}.

\section{Discussion}

We have obtained in this paper the general global stationary cylindrically
symmetric solution for the gravitational field of two identical, rotating
and counter-moving dust beams. Three representations of the interior
solution have been given. They depend on the free parameter $a_0$ and the
arbitrary function $u\left( x\right) $, $u^{\prime }\left( u\right) $ or $%
g\left( u\right) $, which satisfies the condition (25), $k_u>2$ or (19)
respectively. A particular solution with $a_0=1$ and depending on the
arbitrary parameter $a$ has been found by Kramer \cite{seven}. Many of its
nice properties are shared also by the general solution. It satisfies the
dominant energy condition. The energy density of the beams is non-negative.
The axis is regular and elementary flat. The solution is necessarily
non-diagonal. We have studied its stationary alternative. Rotation
compensates gravitational attraction and prevents collapse and appearance of
singularities.

It was shown that two non-zero components of the dust four-velocities are
enough for a solution with arbitrary density profile. In the case of
colliding null-dust three such components are necessary \cite{thirteen}.

The interior solution can be matched at any distance $x_0$ to an exterior
vacuum stationary solution, the Lewis solution \cite{nine,fourteen,fifteen}.
Thus a global solution is formed. An important property is the traditional
appearance of CTC in rotating cylindrically symmetric solutions. This
happens both in the interior when $x_0$ is big enough and in the exterior,
when the Lewis class is induced. The Weyl class is locally equivalent to the
Levi-Civita solution \cite{fourteen}. It is causal and serves as an exterior
for the general static solution.

\end{document}